\documentclass[useAMS]{emulateapj}
\usepackage[]{natbib}

\def\lsim{\lower.5ex\hbox{$\; \buildrel < \over \sim \;$}}
\def\gsim{\lower.5ex\hbox{$\; \buildrel > \over \sim \;$}}

\shorttitle{Detection of GRB~090618 with $RT$-2 Experiment}
\shortauthors{Rao et al.}

\begin{document}
\title{Detection of GRB~090618 with $RT$-2 Experiment Onboard the 
$Coronas-Photon$ Satellite}

\author{A. R. Rao\altaffilmark{1}, J. P. Malkar\altaffilmark{1}, 
M. K. Hingar\altaffilmark{1}, V. K. Agrawal\altaffilmark{1,2}, 
S. K. Chakrabarti\altaffilmark{3,4}, A. Nandi\altaffilmark{2,4}, 
D. Debnath\altaffilmark{4}, T. B. Kotoch\altaffilmark{4}, 
R. Sarkar\altaffilmark{4}, T. R. Chidambaram\altaffilmark{5}, 
P. Vinod\altaffilmark{5}, S. Sreekumar\altaffilmark{5}, 
Y. D. Kotov\altaffilmark{6}, A. S. Buslov\altaffilmark{6}, 
V. N. Yurov\altaffilmark{6}, V. G. Tyshkevich\altaffilmark{6}, 
A. I. Arkhangelskij\altaffilmark{6}, R. A. Zyatkov\altaffilmark{6},
Sachindra Naik\altaffilmark{7}}

\altaffiltext{1}{Tata Institute of Fundamental Research, Mumbai - 400005, 
India. $arrao@mailhost.tifr.res.in$} 
\altaffiltext{2}{Space Science Division, ISRO-HQ, New Bell Road, Bangalore - 560231, India}
\altaffiltext{3}{S. N. Bose National Center for Basic Sciences, Salt Lake, Kolkata - 700098, India}
\altaffiltext{4}{Indian Center for Space Physics, Garia, Kolkata - 700084, India}
\altaffiltext{5}{Vikram Sarabhai Space Center, VRC, Thiruvananthapuram - 695022, India}
\altaffiltext{6}{Moscow Engineering Physics Institute, Moscow - 115409, Russia}
\altaffiltext{7}{Astronomy \& Astrophysics Division, Physical Research Laboratory, Ahmedabad - 380009, India}

\begin{abstract}
We present the results of an analysis of the prompt gamma-ray emission
from GRB~090618 using the $RT$-2 Experiment onboard the $Coronas-Photon$ 
satellite. GRB~090618 shows multiple peaks and a detailed study of the 
temporal structure as a function of energy is carried out. As the GRB was 
incident at an angle of {77$^\circ$} to the detector axis, we have 
generated appropriate response functions of the detectors to derive 
the spectrum of this GRB. We have augmented these results using 
the publicly available data from the $Swift$ BAT detector and show that
a combined spectral analysis can measure the spectral parameters quite
accurately. We also attempt a spectral and timing analysis of individual 
peaks and find evidence for a systematic change in the pulse
emission characteristics for the successive pulses. In particular, 
we find that the peak energy of the spectrum, $E_p$, is found to 
monotonically decrease with time, for the successive pulses of this GRB.
\end{abstract}

    
\section{Introduction}

Gamma-ray bursts (GRBs) are very fascinating cosmic objects in the universe.
Since its discovery in 1973 (Klebesadel et al. 1973), GRBs opened up a new
domain of astrophysical research due to the rich observational characteristics
of the afterglows in a vast range of electromagnetic spectrum from 
$\gamma$-rays to radio wavelengths (see Gehrels et al. 2009 for a review). 
There is a general consensus in the literature that the diverse observational
characteristics are due to the interaction of relativistic matter with the
surrounding medium. The nature and energy source for this relativistic matter,
particularly in the context of the long GRBs, could be in the nature of bulk 
motion of ions (the Fire Ball Model - see e.g., Piran 2004), cannon balls 
emitted from a compact object newly formed in a supernova explosion (Dar 
2006), or particles accelerated by magnetized winds (the electro-magnetic 
model - see Lyutikov 2006 and references therein). The long GRBs are 
associated with supernovae and it is believed that relativistic matter 
of very high bulk Lorentz factor is generated in a conical jet during the 
collapse of a massive star at cosmological distances (see for example, 
Meszaros 2002). The prompt gamma-ray emission has several characteristic 
correlations like the peak energy $E_p$ against the isotropic luminosity 
$E_{iso}$ (Amati et al. 2002), spectral lag against the peak luminosity 
(Norris et al. 2000) etc. (see Gehrels et al. 2009, for a summary of such 
correlations) and these relations are used even in predicting the red shift 
of long duration gamma-ray bursts, although with a large uncertainty (close 
to a factor of 2, see, for example Xiao \& Schaefer 2009). A detailed 
understanding of the prompt emission is necessary to put these correlations 
in a firm footing so that GRBs can be used as cosmological candles and also 
to have a clear understanding of the central engine and the basic jet/cannon-ball 
emission mechanism.

The morphology or temporal profile of the GRBs during the prompt emission 
varies asymmetrically with no apparent structure among the bursts. Some 
GRBs show multiple pulses and the individual pulses in a burst is a separate
and unique emission with varying amplitude and intensity. In the frame work 
of Fireball Model (see, Zhang 2007 and Meszaros 2002 for reviews) these 
pulses are created with different shock strengths at different locations 
of the jet. Observations suggest that in most of the bursts, an individual 
pulse profile is in the shape of fast-rise-exponential-decay (FRED), with 
the width decreasing with energy (Fenimore et al. 1995; Norris et al. 2005).
Spectral lag is another spectro-temporal property which is crucial to
understand the dynamics and energetics of GRBs and can constrain the GRB
models (Ioka \& Nakamura 2001; Shen et al. 2005; Lu et al. 2006). 

GRB~090618 is a very interesting object for several reasons.
It is bright and relatively nearby (redshift $\sim$ 0.5) making it a
good candidate to expect to have a visible supernova (if it is like 
SN1998bw - see Dado \& Dar 2010), though no supernova has yet been associated 
with this GRB. Further, its intense hard X-ray and gamma-ray emission during 
the prompt phase enables one to make a time resolved spectral analysis (see for 
example, Ghirlanda et al. 2010). In this paper, we make a detailed analysis 
of the prompt emission in a wide band X-ray and gamma-ray region using data 
from Swift/ BAT and the $RT$-2 Experiment onboard the $Coronas-Photon$ 
satellite (preliminary results are given in Rao et al. 2009). Since this is 
the first result from this experiment, we describe in detail the methodology 
used in deriving the response matrix and spectral fitting. We augment our 
results by using the publicly available $Swift$ BAT data and make a combined 
spectral fit. We examine the spectral and temporal characteristics of the 
individual pulses during the prompt emission of this GRB and investigate the 
implications to the source emission mechanisms. In \S 2, a summary of 
observations on GRB~090618 is given and in \S 3 a brief description of the 
$RT$-2 Experiment is given. Observations and analysis results ($RT$-2 and 
BAT data) are given in \S 4 and finally in \S 5, a discussion of the results 
are presented along with relevant conclusions.

\section{GRB~090618}

The bright and long gamma-ray burst GRB~090618 was discovered with the 
$Swift$ Burst Alert Telescope (BAT) on 2009 June 18 at 08:28:29 UT (Schady 
et al. 2009a; 2009b; 2009c) at a red shift of $z$=0.54 (Cenko et al. 2009b). 
The GRB was detected by various observatories in X-ray and gamma-ray energies 
such as AGILE (Longo et al. 2009), Fermi GBM (McBreen et al. 2009), Suzaku 
WAM (Kono et al. 2009), Konus-Wind onboard Wind satellite and Konus-RF 
onboard Coronas-Photon satellite (Golenetskii et al. 2009), $RT$-2 Experiment
onboard Coronas-Photon satellite (Rao et al. 2009) etc. The optical afterglow 
of GRB~090618 was soon detected with the Katzman Automatic Imaging Telescope 
(KAIT) (Perley et al. 2009), ROTSE-IIIb (Rujopakarn et al. 2009), Palomar 
60-inch telescope (Cenko et al. 2009a) and various other optical, infra-red 
and radio observatories.

The X-ray afterglow of GRB~090618, as measured by the $Swift$ XRT (Schady et 
al. 2009c), was very bright in X-rays, initially. Soon after, the flux decayed 
rapidly with a slope of $\sim$6 before breaking at T$_0$+310 s (T$_0$ = 08:28:29 
UT) to a shallower slope of 0.71$\pm$0.02 (Beardmore et al. 2009). Further 
breaks at longer time scales were also reported (Schady et al. 2009c). Spectral 
fitting to the $Swift$ data in the range of T$_0$+250 s and T$_0$+1065 s with a 
power law model modified by interstellar absorption yielded a photon index of 
$\sim$2 and the intrinsic absorption of 1.78$\times$10$^{21}$ cm$^{-2}$ 
(Beardmore et al. 2009). They estimated the 0.3-10 keV unabsorbed flux to be 
1.16$\times$10$^{-9}$ erg cm$^{-2}$ s$^{-1}$. 

Significant spectral evolution was observed during the prompt emission of
the burst. Time-averaged spectrum from T$_0$-4.4 to T$_0$+213.6 s was well 
fitted by a power law with an exponential cut-off with photon index of 1.42 
and $E_p$ of 134 keV (Sakamoto et al. 2009). Time-integrated 20 keV -- 2 MeV 
spectra obtained from the Konus-Wind (from T$_0$ to T$_0$+142 s; T$_0$ = 
08:28:24.974 UT) onboard Wind satellite and Konus-RF (from T$_0$ to T$_0$+142 s; 
T$_0$ = 08:28:27.060 UT) instrument onboard Coronas-Photon satellite when fitted 
by GRB (Band) model, provided the values of the low-energy photon index ($\alpha$), 
high-energy photon index ($\beta$) and peak energy ($E_p$) to be -1.28, -2.66, 186 
keV (for Konus-Wind) and -1.28, -3.06, 220 keV (for Konus-RF), respectively
(Golenetskii et al. 2009). The BAT light curve of the GRB was found to be of a 
multi-peak structure with a duration of about 130 s. The time-averaged BAT 
spectrum from T$_0$-5 to T$_0$+109 s can be described by simple power-law model 
with index $\sim$ 1.7 (Baumgartner et al. 2009). The fluence in the 15-150 keV 
band is 1.06 $\pm$0.01 $\times$ 10$^{-4}$ erg cm$^{-2}$. The multi-peaked profile 
was also seen in the 50 keV -- 5 MeV range light curve obtained from the Suzaku 
Wide-band All-sky Monitor (Kono et al. 2009). 

Ukwatta et al. (2010) derived the spectral lag of this GRB using the 
Swift BAT data and found that it supports the existence of a lag-luminosity
relation. Ghirlanda et al. (2010) investigated the time resolved spectral
characteristics of several GRBs using Fermi data and concluded that the 
$E_p$ -- $L_{iso}$ relation holds good during the rising and decaying phases 
of pulses for a few GRBs, particularly for GRB~090618.

\section{$RT$-2 Experiment onboard the Coronas-Photon satellite}

The $RT$-2 Experiment (RT - Roentgen Telescope), which is a part of Indo-Russian
collaborative project of Coronas-Photon mission (Kotov et al. 2008; Nandi et 
al. 2009), is designed and developed for the study of solar hard X-rays in 
$\sim$15 keV to 100 keV energy range. This experiment consists of three instruments
(two phoswich detectors called $RT$-2/S and $RT$-2/G, and one solid-state
imaging detector $RT$-2/CZT) and one processing electronic device ($RT$-2/E).
The $RT$-2/S and $RT$-2/G detector assembly consist of NaI(Tl) / CsI(Na)
scintillators in phoswich assembly viewed by a photomultiplier tube (PMT).
Both the detector assemblies sit behind respective  mechanical slat collimators 
surrounded by a uniform shield of Tantalum material and having different 
viewing angles of 4$^{\circ}$$\times$4$^{\circ}$ ($RT$-2/S) and 
6$^{\circ}\times6^{\circ}$ ($RT$-2/G). The $RT$-2/S covers $\sim$15 keV to 100 keV, 
extendable up to 1 MeV, whereas the use of an Aluminum (Al) filter in $RT$-2/G sets 
the lower threshold at $\sim$ 20 keV. The $RT$-2/CZT consists of three CZT detector
modules (OMS40G256) and one CMOS detector (RadEye-1) arranged in a 2$\times$2
array. Each CZT module consists of 256 individual detectors (pixel dimension 
of 2.5 mm $\times$ 2.5 mm), which are controlled by 2 ASICs. The CMOS detector
consists of 512$\times$512 pixels of individual pixel dimension of 48 \micron.
The entire CZT-CMOS detector assembly is mounted behind a collimator with two
different types of coding devices, namely coded aperture mask (CAM) and Fresnel
zone plate (FZP), surrounded by a uniform shield of Tantalum material and has
varying viewing angles of 6$'$--6$^{\circ}$ depending on different configurations
of the collimator (Nandi et al. 2010). The $RT$-2/CZT payload is the only imaging 
device in the Coronas-Photon mission to image the solar flares in hard X-rays in 
the energy range of 20 to 150 keV. All three-detector systems are interfaced with 
the satellite system (called SSRNI) through $RT$-2 Electronic processing payload 
($RT$-2/E). The $RT$-2/E receives necessary commands from the satellite 
system and passes it to the individual detector system for proper functionality
of the detector units and acquires data from the detector system and stores 
in its memory for further processing. 

The mission was successfully launched from Plesetsk Cosmodrome, Russia on 2009 
January 30. To maximize the Sun observation time, the satellite was put into a 
low earth (500 km) Sun-synchronous near-polar (inclination 82.5$^\circ$) orbit. 
The test and evaluation results of this payload are described in Debnath et al. 
(2010), Kotoch et al. (2010), Nandi et al. (2010), Sarkar et al. (2010) and 
Sreekumar et al. (2010). Some details of the experiment can also be found at 
http://csp.res.in/rt2-main.html.

\section{Observation and Analysis}

\subsection{$RT$-2 Observations}

During the GRB event, the $RT$-2 payload was completely in the `SHADOW' 
mode (away from the Sun) which started at 08:16:10.207 UT and ended at
08:37:35.465 UT. GRB~090618 was detected by the $RT$-2 instruments 
(Rao et al. 2009) with a large off-axis angle of $77^\circ$. Data from 
$RT$-2/S and $RT$-2/G are used for the present analysis. The scientific 
data from the detectors were stored in the memory of $RT$-2/E and then 
compressed using the onboard software before being transferred to the 
satellite system (known as SSRNI) for down-link to the ground station. 
During the `SHADOW' mode, the spectra are accumulated every 100 s (see 
the spectral analysis section) and eight channel count rates (for each 
detector) are accumulated every 1 sec. These eight channels are divided 
equally between the 15 -- 100 keV data from the 3 mm thick NaI(Tl) 
detector and the 26 -- 1000 keV data from the 25 mm thick  CsI(Na) 
detector, operated in a Phoswich mode. For the present work, we use the 
latter four channel data, which have the ranges of 26 -- 59 keV, 
59 -- 215 keV, 215 -- 330 keV, and 330 -- 1000 keV, respectively.

The `SHADOW' mode data of $RT$-2/S and $RT$-2/G detectors of `GOOD' time 
(away from high background regions) span of $\sim 800$ s from 08:23:27 UT 
to 08:36:47 UT were analyzed. During this time, the satellite was completely 
away from both the polar cap and high background SAA regions. GRB~090618 
was detected by both the scintillator detectors. The light curves of 1 s 
time resolution were generated from the data obtained from both the 
detectors at different energy bands. Spectral data of 100 s, during the 
present GRB event, were also analyzed to study the evolution of the GRB 
spectrum.

\subsection{Swift BAT Observations}

Burst Alert Telescope (Barthelmy et al. 2005) onboard the {\it Swift} mission
(Gehrels et al. 2004) is a highly sensitive, large field of view instrument,
primarily designed to monitor the sky to detect gamma-ray events. It consists 
of an array of CdZnTe detectors, located behind a coded aperture mask. 
{\it Swift} also has an X-ray Telescope (XRT; Burrows et al. 2005) and 
UV/Optical Telescope (UVOT; Roming et al. 2005) which can make follow up
observations within a few hundred seconds after the trigger.

BAT registered GRB~090618, triggered  at 08:28:29 UT. The BAT position of the
burst with 3$\arcmin$ uncertainty is RA(J2000), Dec(J2000) = 294.021, +78.353
(Schady et al. 2009a). Astrometrically corrected position using Swift-XRT data
and Swift-UVOT data is RA(J2000), Dec(J2000) = 293.99465, +78.35677 (Evans et 
al. 2009; Schady et al. 2009c). We have analyzed the Swift BAT data to understand 
the prompt emission from this GRB.

We used {\bf heasoft6.5.1} for our analysis. First we created the detector 
plane image using the task {\it batbinevt} and then the detector mask file 
was created using task {\it batdetmask}. The quality map file was created 
using task {\it bathotpix}. Mask weighting using new XRT and UVOT position 
to BAT data was applied using task {\it batmaskwtevt}. Then mask weighted 
light curves were generated using task {\it batbinevt}. The GRB light curve 
shows multi-peaked structure with a precursor and the main burst starts at 
T$_0$+50 s. We generated light curves for the full burst (T$_0$ to T$_0$+180) 
in 4 energy bands: 15-25 keV, 25-50 keV, 50-100 keV, 100-200 keV
(these are the energy ranges used in Ukwatta et al. 2010, for timing analysis).

\subsection{Timing Analysis}

The GRB was detected in the wide band of 26 keV to 1000 keV with both the
$RT$-2 detectors. Since the GRB is incident at a large incident angle,
meaningful light curves were available from the CsI(Tl) detectors in different 
energy bands (26-59 keV, 59-215 keV, and 215-1000 keV ranges for $RT$-2/S 
and 26-59 keV, 59-215 keV, 215-330 keV, and 330-1000 keV ranges for $RT$-2/G). 
The light curves, obtained from $RT$-2 and BAT instruments, are shown in Figure~1 
with a bin time of 1 s and and the time is given with respect to the BAT trigger 
time T$_0$ of 2009 June 18 08:28:29 UT. The BAT light curves are in the units of 
mask-weighted counts, which are essentially background subtracted counts
per fully illuminated detector for an equivalent on-axis source. The $RT$-2 
count rates are the combined light curves from $RT$-2/S and $RT$-2/G detectors 
and they are normalized to the nominal detector area of 100 cm$^2$ for each 
detector. The light curves are arranged with increasing energy (from the top) 
and wherever possible light curves of comparable energy are plotted together 
(25 -- 50 keV BAT data and 26 -- 59 keV $RT$-2 data in panel b and 100 -- 200 
keV BAT data and 59 -- 215 keV $RT$-2 data in panel c). For clarity, the 
59 -- 215 keV light curve from $RT$-2 is scaled down by a factor of 10 and 
vertically shifted up by 0.1 count s$^{-1}$.  

The burst profile was found to be having a complex structure. The entire burst 
episode lasted for about 150 s with the brightest pulse detected at T$_0+65$ s. 
The other two weak emissions are registered at T$_0+85$ s and at T$_0+115$ s with 
low intensity. Another instrument, Konus-RF onboard the Coronas-Photon satellite, 
also detected the GRB~090618 with an identical burst profile (Golenetskii et al. 
2009). The burst profile is not clearly detected in the lowest energy band of 
$RT$-2 (26 - 59 keV), whereas in the high energy X-rays, burst is detected with
significant emission. There is a very significant softening of the spectrum in the 
precursor with the 25 -- 50 keV light curve peaking several seconds after the high 
energy light curves. The main pulse at T$_0$+65 s shows a double structure, the 
50 -- 100 keV BAT profile showing a close similarity to the 59 -- 215 keV $RT$-2 
light curve. The pulse at T$_0$+85 s shows a single smooth structure in the 
100 -- 200 keV (similar to the 59 -- 215 keV $RT$-2 light curve), with an indication 
of multiple structures in the low energies.

We have attempted to model the pulse with FRED profile developed by Kocevski et al. 
(2003). The empirical relation for the flux (photon counts s$^{-1}$) distribution 
is given by,

$$
F(t) = F_m(\frac {t}{t_m})^r[\frac {d}{d+r} + \frac {r}{d+r}
	(\frac {t}{t_m})^{(r+1)}]^{-(r+d)/(r+1)},
\eqno (1)
$$
where, $F_m$ is the maximum flux at $t_m$, $r$ and $d$ are the rising and
decaying indices, respectively. First we fit the total light curves
(26 -- 1000 keV for $RT$-2 and 15 -- 200 keV for BAT). The start times for 
fitting (for each pulse) were kept fixed at $t_m - t_{st}$, where $t_{st}$ 
were varied between 15 s and 25 s. Since there were negligible changes in 
the derived parameters, particularly for the measurement of width, we kept 
$t_{st}$ fixed at 25 s. The reduced $\chi^2$ for the $RT$-2 data is 1.38 (for 
75 dof). For the higher sensitivity BAT data, however, the FRED profile
does not take into account the sub-structures in the pulses, and hence 
formally unacceptable fits (reduced $\chi^2$ $\ge$ 10), were obtained. 
Since our main emphasis is to find the broad pulse characteristics, we give 
below the parameters obtained from a FRED fitting. Further, we find that the 
derived parameters for the pulses 1 \& 2 from the BAT data (these two pulses 
have a large overlap) are quite sensitive to the initial parameters used
for fitting and hence the quoted formal errors could be underestimates.

The resultant derived parameters, along with the nominal 1-$\sigma$ errors 
(obtained from the criterion $\Delta \chi^2$ = 2.7), are given in Table 1.
There is a reasonable agreement in the derived parameters, obtained from 
$RT$-2 and BAT data. While fitting, an upper bound of 1000 is kept for the
parameter $d$ (for $d >> r$, F(t) becomes independent of $d$). We have 
measured the pulse width as the FWHM value of the fitted light curve. The 
fitted light curves with burst profiles are shown in Figure~2 for the total 
energy band of 26 to 1000 keV for the $RT$-2 detectors (bottom panel) and 
for the total energy band of 15 -- 200 keV for the BAT detector (top panel). 
The fitting procedure is repeated for light curves of different energies. 
The times of the maximum emission $t_m$, however, were kept fixed at those 
values obtained from the fitting for the full energy light curves, of the 
respective detectors ($RT$-2 and BAT). The value of  $t_m$ could depend on 
energy upto about a second for these energies (see the delay analysis presented 
later), which might result in a further systematic error in the width measurement 
of $<$ 1 s. We have plotted the width as function of mean energy in Figure~3. Mean 
energies are  calculated as the mean energy of incident photons in the detector, 
based on the response function of the detectors, and they are derived to 
be 125.62 keV, 266.6 keV, and  427.5 keV, for the 59 -- 215 keV, 215 -- 330 
keV, and 330 -- 1000 keV bands, respectively for the $RT$-2 data.
The mean energies for the four energy bands of BAT data are 20.82 keV,
35.45 keV, 68.07 keV, and 123.73 keV, respectively.

The first two pulses have similar profiles in both data sets, though 
in the $RT$-2 data the first pulse is much stronger than the second. There is 
an indication of further sub-structures in the BAT data which is not very
apparent in the $RT$-2 data, possibly due to the lower sensitivity. The width
decreased monotonically with energy. Also, there is a trend of steepening of 
this trend for the latter pulses, particularly for the third pulse. 
For example, by defining a width index $\xi$ such that $width \propto E^{-\xi}$ 
(Fenimore et al. 1995), we find $\xi$ to be 0.18, 0.07, 0.14 and 0.05, respectively 
for the four pulses, for a combined fit to the $RT$-2 and BAT data. These parameters 
are summarized in Table~3. 

To investigate the softening of the spectra, we have performed cross correlation 
analysis between the light curves of various energies using the BAT data. We have 
taken the 15 -- 25 keV  as the base energy. Cross correlation is done for the full 
light curves as well as in parts: the first part includes the precursor (T$_0$ to 
T$_0$+50), the second part includes pulse 1 and 2 (T$_0$+50 to T$_0$+77) and part 3 
and 4 include pulses 3 and 4, respectively. We have used the utility {\it crosscor} 
of the XRONOS software package (a part of the HEASOFT software package of HEASARC)
to derive the cross correlation function (CCF) with respect to lag. A Gaussian 
function was used to fit the CCF and measure the time lag and the error is estimated 
using the criteria $\Delta \chi^{2}$=4.0 (see Dasgupta \& Rao 2006). The results 
of cross correlation analysis are given in Table~2. We found soft lags which show 
clear energy dependence. The measured lag increases with energy and they are shown in 
Figure~4. There is also a tendency for this steepening to be flatter for the latter 
pulses. Defining a delay index $d_i$ such that $delay = a + d_i ln (E)$, where $a$ 
is a constant, we find $d_i$ to be very steep for the precursor (-3.71) and it 
increased from -0.50 for Part 2 (pulses 1 \& 2) to -0.13 for Part 3 (pulse 3). For 
the fourth part (pulse 4), however, $d_i$ is found to be -0.92. The values of $\xi$ 
and delay index ($d_i$) are compiled in Table~3.

Ukwatta et al. (2010) have derived  delays for the light curve of T$_0$+46.01 
to T$_0$+135.35 s as -0.171 s, -0.314 s and -0.579 s, respectively, in three
energy bands given in Table~3, which is close to the average values of delays
for part 2 and 3 (T$_0$+50 to T$_0$+100 s) reported in Table~3 (-0.247 s, 
-0.365 s, and -0.630 s, respectively, for the three energy bands). We have
followed a method of cross-correlation analysis similar to that followed in
Ukwatta et al. (2010) and we have confirmed that we get consistent results 
in the time span used in their work.

\subsection{Spectral Analysis}

We have analyzed the spectral data during the GRB event. It showed the 
typical Band spectrum with peak energy at about 164 keV and integrated 
20 keV - 1 MeV fluence of 2.8 $\times$ 10$^{-4}$ ergs cm$^{-2}$. 
We used 15-200 keV BAT data and the 100 -- 650 keV $RT$-2 data of the 
GRB~090618 to perform joint spectral fitting.  

The spectral information was available in the $RT$-2 data, every 100 s.
The output from each detector is passed through two amplifiers of
different gains (G1 and G2) and spectral data were available for each 
of the amplifiers: 1024 channel spectra for G1 (covering the energy 
range of 26 -- 215 keV for the CsI detector) and 256 channel spectra 
for G2 (covering the energy range of 215 -- 1000 keV). The number of 
channels are suitably rebinned for the spectral fitting.

$RT$-2/S and $RT$-2/G detectors are essentially collimated Phoswich 
detectors user for solar flare studies (Debnath et al. 2010). The 
shielding used for these detectors, however, are optimized to use them 
for hard X-ray spectroscopic studies of solar flares in 15 -- 100 keV 
region and as omni-directional hard X-ray/gamma-ray detectors between 
$\sim$ 50 keV to 1000 keV (Sarkar et al. 2010). GRB~090618 is incident 
at an angle of 77$^\circ$, and we have used a Monte Carlo (MC) simulation
technique using the GEANT4 toolkit to derive the spectral response of the 
$RT$-2 detectors for this large incident angle (see Sarkar et al. 2010 
for details). We created a mass model of the detector including all 
parameters like detector sizes, collimator, shield materials and 
mechanical support structures. Incident photons are used for the 
simulation in the energy range of 1 -- 1000 keV in 50 equal bins in 
log scale and 10$^5$ photons in each bin are considered. The number of 
photons registered in the detectors as a function of energy is normalized 
to the incident photons for a GRB model with the spectral 
parameters of $\alpha$ = -0.32, $\beta$ = -5.0 and E$_0$ = 67.7 keV.
The results are insensitive to model parameters (see later). 
This normalization is deemed as the effective area of the detectors.
The derived effective areas for the NaI and CsI detectors are
shown in Figure~5. Since the effective area of CsI is about an order
of magnitude larger than the NaI detector, we have taken only the CsI data
(and the corresponding response function) for our spectral fitting.
The response matrix for the $RT$-2 detectors are generated using 
the $genresp$ tool of $ftools$. The channel-energy conversion is derived 
from the background spectral line at $\sim$58.0 keV (Nandi et al. 2009) and 122 keV 
line due to the onboard calibration source (Co$^{57}$) and the energy resolution 
function is taken from the ground calibrations. Using this response matrix, a joint 
fit to the $RT$-2 and BAT data is performed. From the derived best fit 
spectral parameters, MC simulation is again carried out and the $RT$-2 
response matrix is again created. Convergent results were obtained in 
the first trial itself, indicating that direct blocking from the 
surrounding material and interaction with detector play a major 
part of the response and the second order effects like scattering are
unimportant for response matrix generation, at least to the level of 
sensitivity achieved for GRB~090618.

Except for a 2 mm aluminum filter in $RT$-2/G, both the detectors are 
identical in performance. The setting for deriving the onboard 
counts in the high energy channels is lower in $RT$-2/S and hence we
have a single energy bin above 215 keV in $RT$-2/S (compared to the two 
energy bins in $RT$-2/G). Though we get consistent spectral results for 
both the detectors, we report the spectral parameters derived from 
$RT$-2/S which has better spectral response above 100 keV. While making 
a simultaneous fit to the BAT data, we kept the relative area between 
$RT$-2 and BAT as a free parameter and it is derived to be $1.30$ for the 
combined fit to the full data. The relative area between $RT$-2 G1 and
G2, however, is kept fixed at what is dictated by the response matrix. 
For the time resolved spectral fit the relative area between $RT$-2 and 
BAT is kept fixed at the value obtained for the full data (i.e., $1.30$). 

The spectrum can be well fitted with the model introduced by Band et al. 
(1993). In this model, two power-laws are smoothly joined at 
$(\alpha-\beta)$E$_0$, where $\alpha$ is the first power-law index 
and $\beta$ is the second power-law index, and E$_0$ is the break energy. 
The best fit spectral parameters are given in Table~3, along with the calculated 
peak energy E$_p$ (= E$_0$ (2 + $\alpha$)) and the best fit $\chi^2$ and the degrees 
of freedom used for the fitting. Figure~6 shows the best fit model along with the 
unfolded spectrum. We divided the BAT data in 4 parts and for each part 
we perform the spectral analysis. Since $RT$-2 does not have time-resolved
spectral data, we have used the count rates as broad spectral data for this 
time-resolved spectral analysis. The results are shown in Table~3. Timing
analysis parameters (width index $\xi$  and delay index $d_i$) are also given 
in Table~3. We note that the spectral results agree quite well with that
obtained from the $fermi$ data (Ghirlanda et al. 2010). The values of 
E$_p$, derived in the present work, are 264 keV, 248 keV, 129 keV, and 
33  keV respectively for the four parts which compares well with the value 
of 296 keV, 319 keV, 180 keV, and 81 keV derived for the peak of the light
curves in these regions in the $fermi$ data (time ranges, after T$_0$, of 
3 -- 14 s, 63 -- 67 s, 80 -- 85 s, and 114 -- 130 s, respectively). 

\section{Discussion and conclusions}

In this paper, we have presented a method to measure the spectral parameters 
of the prompt emission using the recently launched $RT$-2 detectors and the 
Swift-BAT. Though $RT$-2 has been made to primarily study solar activities, 
we find that above $\sim$ 50 keV, $RT$-2 essentially acts as an all sky hard 
X-ray monitor. Thus, it can be used to measure the spectral and timing 
characteristics of the prompt emission of GRBs.

GRB~090618 shows multiple peaks. It shows a systematic softening of the
spectrum for the successive pulses which is associated with the variations 
in the timing parameters. For the successive peaks in the GRB, the peak 
energy shifts to lower values, the width of the pulse varies sharply with 
energy and the delay (which is lower for the latter pulses) as a function 
of energy shows a flatter dependence on energy. The parameter $\xi$, 
characterizing the dependence of the pulse width with energy, also shows 
a strong pulse (and hence time) dependence.

Fenimore et al. (1995) used the width of individual pulses in several GRBs 
detected by BATSE and showed that the dependence on energy is a power-law, 
with an index of $\sim$0.4. Borgonovo et al. (2007) measured the width of 
several GRBs of known redshifts and found that $\xi$ shows a continuous 
distribution. A large number of bursts in their work show $\xi$ to be 
peaking around 0.1 -- 0.2. It appears that GRB~090618 belongs to this 
class of GRB with a narrow distribution of $\xi$. Interestingly, though 
the value of $\xi$ changes from pulse to pulse in this GRB, it is
in the range of 0 -- 0.1.

Various explanations are offered to understand the lags seen in GRBs.
Shen et al. (2005) have estimated the contributions from the
relativistic curvature effect. These effects can explain contributions
to the lag of the order of 10$^{-2}$ -- 10$^{-1}$ s. The lags observed
in GRB~090618 is much larger than this and hence quite unlikely
to be due to the curvature effect. Ioka \& Nakamura (2001) have
computed the kinematic dependence of lag caused by the viewing angle
which could produce the observed dependences. The systematic 
pulse-to-pulse variation of the properties detected in GRB~090618, 
however, would be difficult to understand in this framework. Spectral 
evolution, too, can reproduce some part of the lag (Kocevski \& Liang 
2003; Hafizi \& Mochkovitch 2007). The results from the present observations
support the conclusion of Hakkila et al. (2008) that the spectral lags
are pulse rather than burst properties.  

The individual pulses in a GRB could be either due to emission from
multiple shock locations when a jet material encounters the supernova 
ejecta or due to the weakening of the relativistic matter and/or 
the emission of a fresh relativistic matter from the central
engine. Detection of properties of a GRB pulse quite different 
from the previous pulses has implications for the nature of
the central engine. If the central engine is shown to emit
multiple ejection episodes it will strongly favor a black hole as
the candidate as against a highly magnetized neutron star
(see for example, Metzger 2010 for a discussion on the GRB central
engines). In the case of GRB~090618, there is a marginal evidence
for the last pulse (pulse 4) to be different and could be a candidate
for a separate emission from the central engine. A detailed study
of several such multi-peaked bursts are required to draw a firm
conclusion. For example, discovery of a hard GRB pulse, after a 
X-ray pulse, would certainly favor a black hole as a candidate
for the central engine.

We have calculated the isotropic energy (E$_{iso}$) for this GRB by 
considering a standard cosmology model for a flat universe with q$_0$ 
= 1/2 and H$_0$ = 70 km s$^{-1}$ Mpc$^{-1}$. Using the measured redshift 
of 0.54 and the measured integrated fluence in the energy range of 20 
keV to 1 MeV of 2.8 x 10$^{-4}$ ergs cm$^{-2}$, we calculate E$_{iso}$ to
be 2.21 $\times$ 10$^{53}$ ergs (beaming effects are neglected). The 
measured time-averaged peak energy (E$_p$) for the entire burst is around
164 keV, which gives the intrinsic peak energy (E$_{p,i}$) of 252 keV. Based 
on the measured values of E$_{p,i}$ and E$_{iso}$, it is found that the 
GRB~090618 closely follows the `Amati' relation with a minor deviation, 
which is within the 2$\sigma$ scatter. Hence, it could be concluded that 
GRB~090618 is a standard candle for the category of long duration GRBs 
alongwith various intrinsic properties that are discussed in this paper.

The recent detection of polarization in GRB~090102 (Steele et al. 2009)
indicates the presence of ordered magnetic field in the source of GRBs 
during the prompt emission. The present measurement of the spectral and 
temporal parameters of GRB~090618 shows that the individual pulses show 
distinct behaviors. 

\begin{acknowledgements}
This work was made possible in part from a grant from Indian Space Research 
Organization (ISRO). The whole-hearted support from G. Madhavan Nair, 
Ex-Chairman, ISRO, who initiated the $RT$-2 project, is gratefully 
acknowledged. Significant contributions from several organizations 
for the realization of the $RT$-2 payload is gratefully acknowledged.
This research has made use of data obtained through HEASARC Online 
Service, provided by the NASA/GSFC, in support of NASA High Energy 
Astrophysics Programs. We thank the anonymous referee of this paper for the
very detailed comments which helped in improving the quality of the 
paper.
\end{acknowledgements}

\begin{figure}
\centering
\includegraphics[angle=-90,scale=.60]{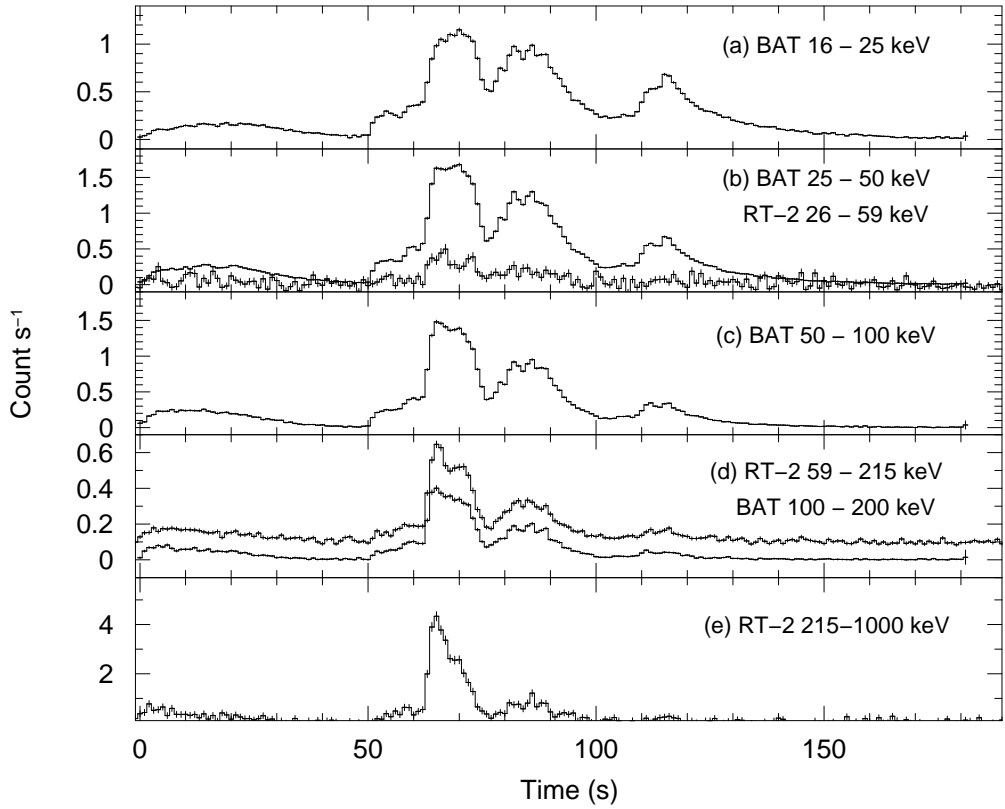}
\caption{The light curves of GRB~090618 obtained from Swift/ BAT and the
$RT$-2 experiments, with a bin time of 1 s. The BAT light curves are in 
the units of mask-weighted counts (see text). The $RT$-2 count rates are 
the combined light curves from $RT$-2/S and $RT$-2/G normalized to the 
nominal detector area of 100 cm$^2$ for each detector. For clarity, the 
59 -- 215 keV light curve from $RT$-2 is scaled down by a factor of 10 and 
vertically shifted up by 0.1 count s$^{-1}$. The time is given with respect 
to the BAT trigger time of 2009 June 18 08:28:29 UT. }
\label{fig1}
\end{figure}

\begin{figure*}
\plotone{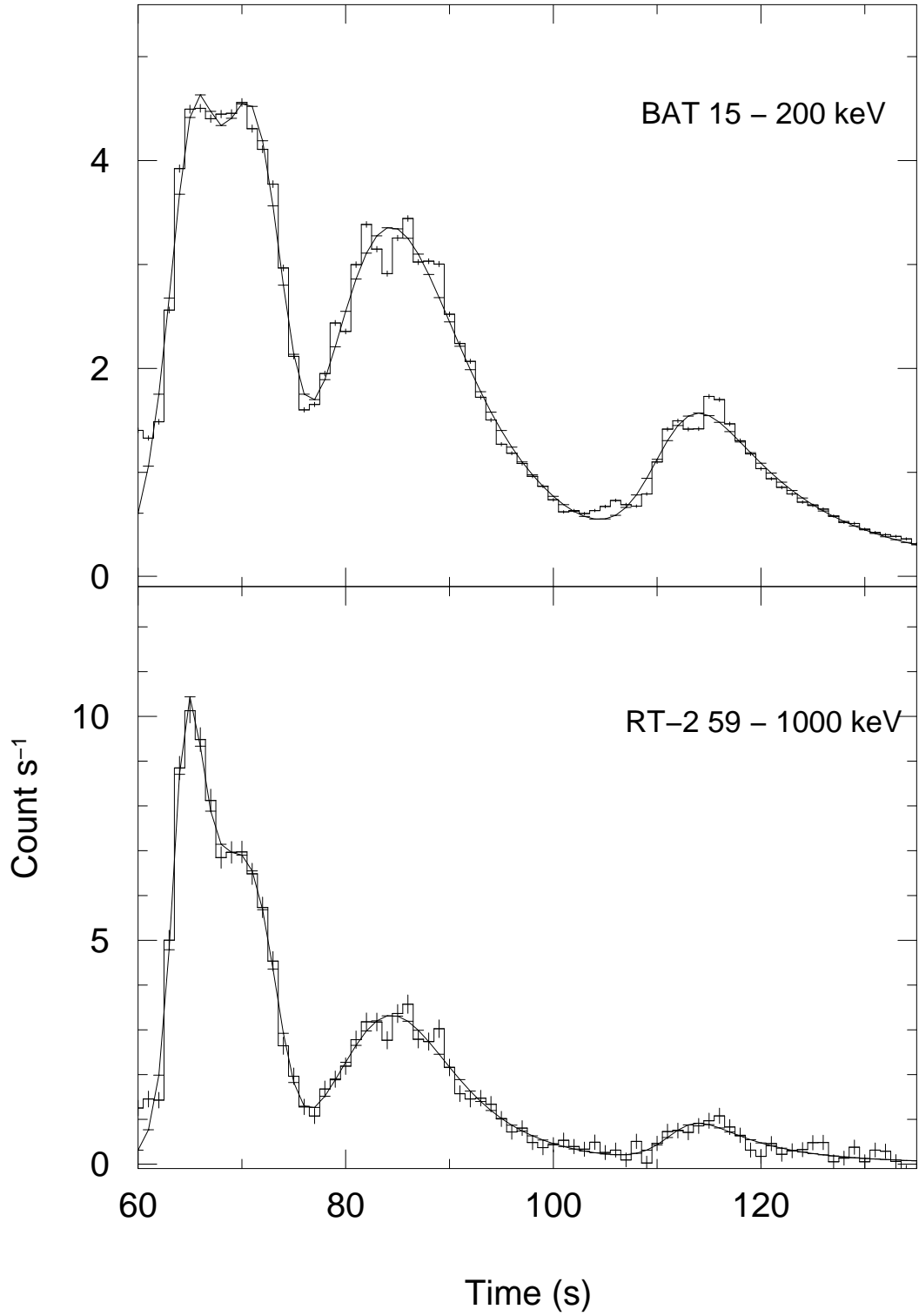}
\caption{The light curves of GRB~090618 obtained from Swift/ BAT and the
$RT$-2 experiments for the respective full energy ranges, shown along with a
model consisting of four FRED profiles each. See Table 1 for the model 
parameters.}
\label{fig2}
\end{figure*}

\begin{figure}
\centering
\includegraphics[angle=-90,scale=.60]{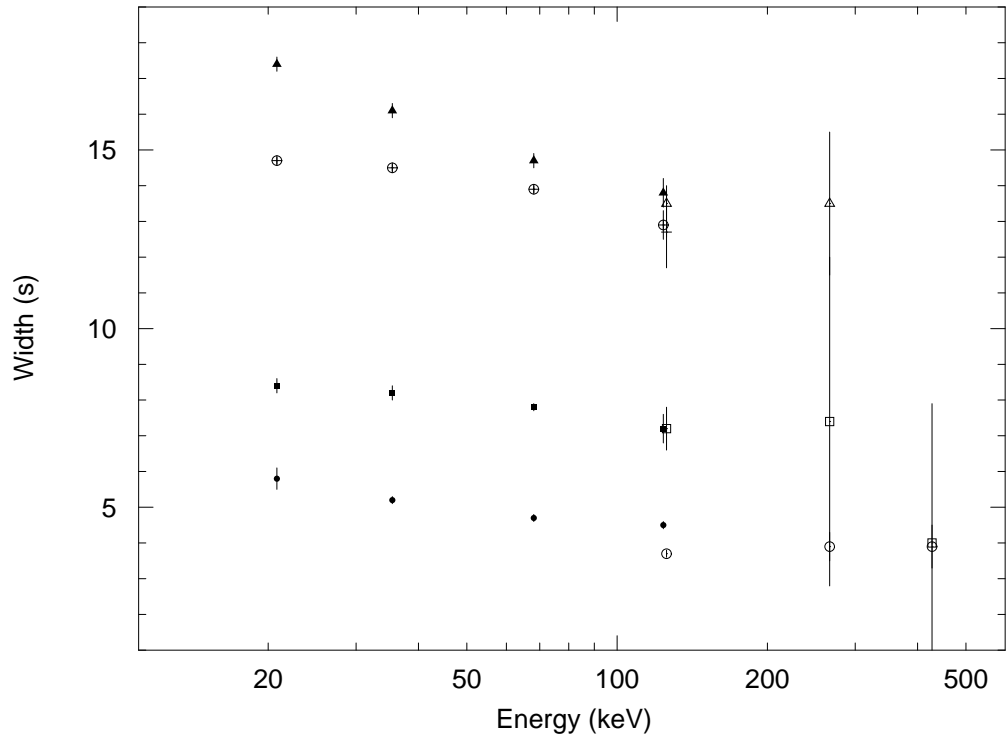}
\caption{The pulse width variation as a function of average energy.
For the four pulses (1 to 4, see text), data are plotted as 
filled circles, filled squares, filled triangles and circles 
with plus signs, respectively, for the BAT data and open circles, 
open squares, open triangles and circles, respectively, for the
$RT$-2/G data.}
\label{fig3}
\end{figure}

\begin{figure}
\centering
\includegraphics[angle=-90,scale=.55]{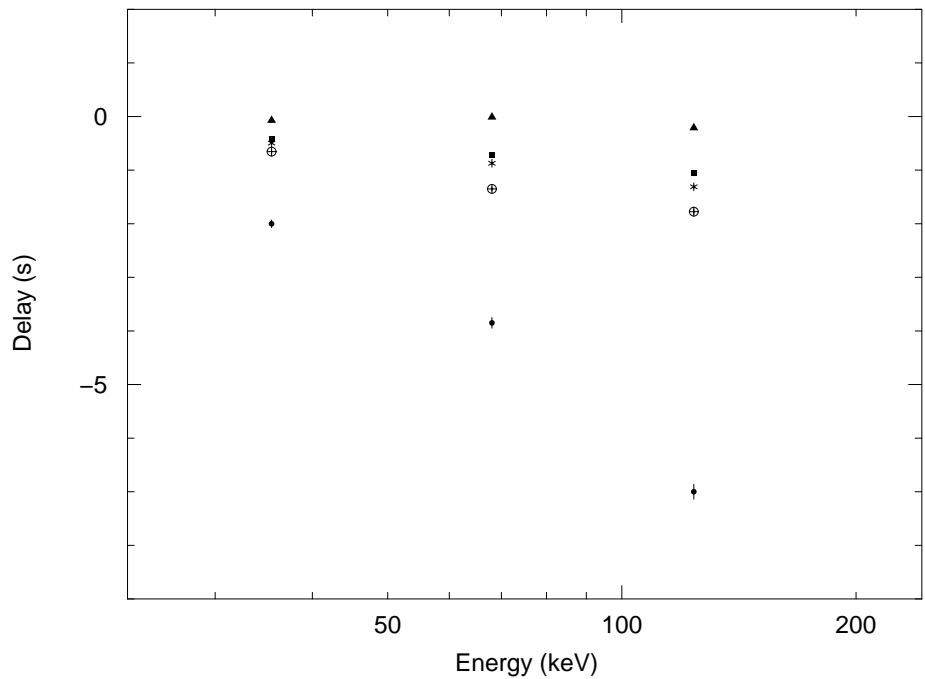}
\caption{The measured delay of light curves with mean energy, E, with 
respect to the 15 -- 25 keV light curve, plotted against E from the $Swift$ 
BAT data. Data for the four parts (see text) and the total light curve are 
shown as filled circles, filled squares, filled triangles, circles with plus
sign, and stars, respectively.}
\label{fig4}
\end{figure}

\begin{figure}
\centering
\includegraphics[angle=-90,scale=.60]{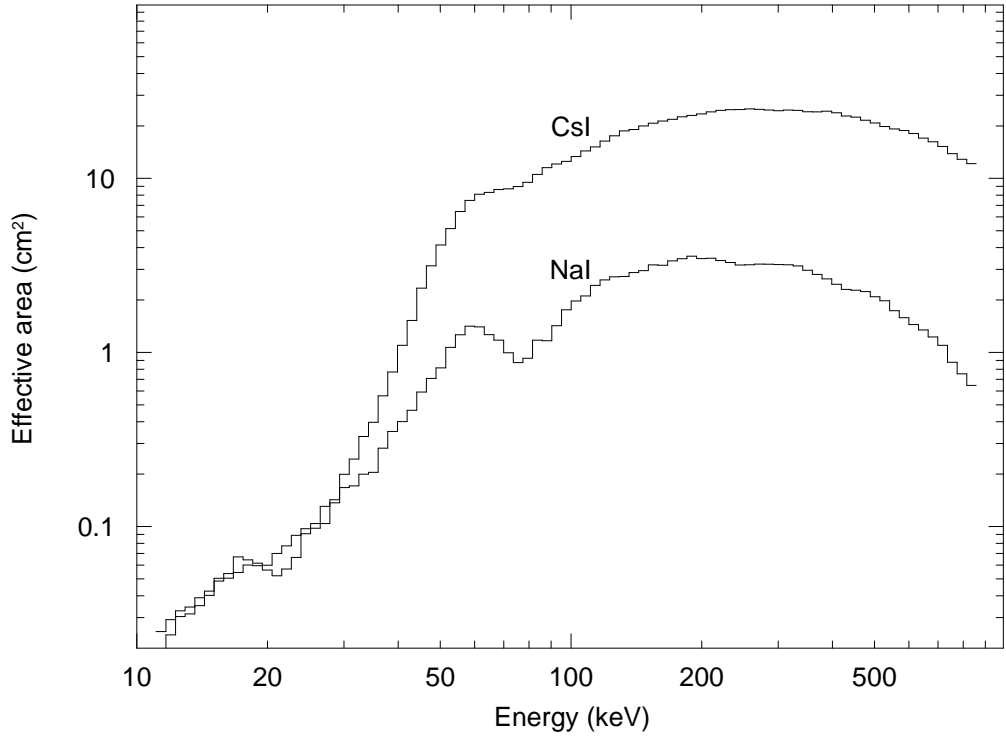}
\caption{The effective area of $RT$-2 NaI(Tl) and CsI(Na) detectors, from a
Monte Carlo simulation for the GRB incident at an angle of 77$^\circ$. }
\label{fig5}
\end{figure}

\begin{figure}
\centering
\includegraphics[angle=-90,scale=.55]{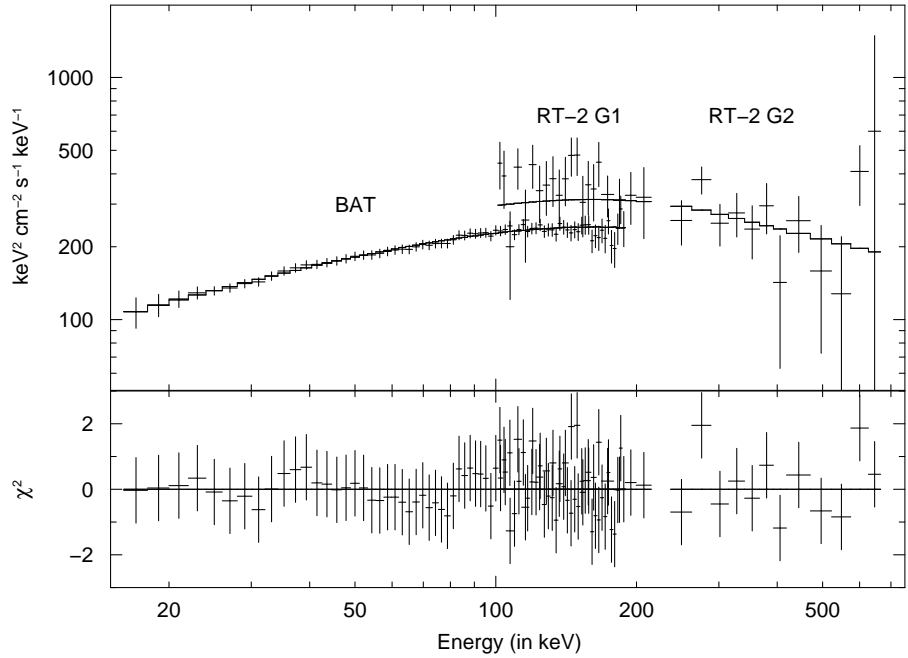}
\caption{The unfolded spectrum (in energy units) of GRB~090618 using $Swift$ BAT and $RT$-2
data, is shown along with the best fit model. The residuals are given in the bottom panel.}
\label{fig6}
\end{figure}

\begin{table}[h]
\caption{Pulse characteristics of GRB~090618}
\label{First Table}
\centering
\begin{tabular}{|c|c|c|c|c|c|c|}
\hline
 Pulse & Energy range (keV) & $F_m$ (s$^{-1}$) & $t_m$ (s) & $r$  & $d$   & Width (s) \\
\hline
1 &26 -- 1000 (RT2) & 8.7$\pm$0.8 &   64.8$\pm$0.2 &   27$\pm$4 &   15$\pm$5 &    3.9$\pm$0.3 \\
 2 &  & 5.9$\pm$0.7 &   70.5$\pm$0.3 &    8$\pm$1 &      1000 &    7.2$\pm$1.1 \\
 3 &  & 3.3$\pm$0.1 &   84.4$\pm$0.2 &    6$\pm$1 &    7$\pm$1 &   13.4$\pm$0.5 \\
 4 &  & 0.85$\pm$0.14 & 114.1$\pm$1.0 & 13$\pm$6 &    5$\pm$2 &    9.7$\pm$0.6 \\
\hline
1 & 59 -- 215 (RT2)   & 4.4$\pm$0.3 &   64.8        &   25$\pm$2 &   20$\pm$5 &    3.7$\pm$0.1 \\
 2 &  & 4.0$\pm$0.2 &   70.5        &    8$\pm$1 &  1000     &    7.2$\pm$0.6 \\
 3 &  & 2.3$\pm$0.1 &   84.4        &    6$\pm$1 &    7$\pm$1 &   13.5$\pm$0.5 \\
 4 &  & 0.6$\pm$0.1 &  114.1        &    8$\pm$3 &    5$\pm$2 &   12.7$\pm$1.0 \\
\hline
1 &215 -- 330 (RT2)  & 2.8$\pm$0.4 &   64.8        &   29$\pm$5 &   14$\pm$6 &    3.9$\pm$0.4 \\
 2 &  & 1.5$\pm$0.4 &   70.5        &    8$\pm$4 &  895$\pm$35 &   7.4$\pm$4.6 \\
 3 &   &0.76$\pm$0.11 & 84.4        &    6$\pm$3 &    8$\pm$6 &   13.5$\pm$2.0 \\
\hline
1 &330 -- 1000 (RT2)     & 1.51$\pm$0.21 &   64.8      &   32$\pm$9 &   12$\pm$1 &    3.9$\pm$0.6 \\
 2 &  & 0.36$\pm$0.22 &   70.5      &   15$\pm$9 & 1000      &    4.0$\pm$3.9 \\
\hline
1 &15 - 200 (BAT)   & 2.80$\pm$0.03 &   65.0$\pm$0.1 & 13.6$\pm$0.4 & 39$\pm$4 &   5.0$\pm$0.1 \\
 2 &  & 4.31$\pm$0.03 &   70.5$\pm$0.1 & 7.0$\pm$0.1 & 1000     &   8.0$\pm$0.1 \\
 3 &  & 3.36$\pm$0.02 &   84.4$\pm$0.1 & 5.1$\pm$0.1 & 5.4$\pm$0.1& 15.9$\pm$0.1 \\
 4 &  & 1.41$\pm$0.01 &  114.3$\pm$0.1 & 8.5$\pm$0.2 & 3.4$\pm$0.1 &14.5$\pm$0.1 \\
\hline
1 &15 -- 25 (BAT)    & 0.52$\pm$0.02 &   65.0      &    9.9$\pm$0.6 &  895$\pm$10 &  5.8$\pm$0.3 \\
 2 &  & 1.09$\pm$0.02 &   70.5      &    6.7$\pm$0.2 &  771$\pm$10 &  8.4$\pm$0.2 \\
 3 &  & 0.97$\pm$0.01 &   84.4      &    4.9$\pm$0.2 &  4.6$\pm$0.2 & 17.4$\pm$0.2 \\
 4 &  & 0.54$\pm$0.01 &  114.3      &   10.4$\pm$0.4 &  2.8$\pm$0.1 & 14.7$\pm$0.1 \\
\hline
1 &25 -- 50 (BAT)    & 0.96$\pm$0.04 &   65.0      &   12.8$\pm$0.8 &   44$\pm$13  &    5.2$\pm$0.1 \\
 2 &   &1.58$\pm$0.03 &   70.5      &    6.8$\pm$0.1 &   1000      &    8.2$\pm$0.2 \\
 3 &  & 1.28$\pm$0.01 &   84.4      &    5.1$\pm$0.2 &  5.3$\pm$0.2 &   16.1$\pm$0.2 \\
 4 &  & 0.54$\pm$0.01 &  114.3      &    8.5$\pm$0.3 &  3.4$\pm$0.1 &   14.5$\pm$0.1 \\
\hline
1 &50 -- 100 (BAT)   & 0.99$\pm$0.03 &   65.0   &   15.5$\pm$0.8 &   30$\pm$6 &    4.7$\pm$ 0.1 \\
 2 &  & 1.31$\pm$0.03 &   70.5   &    7.2$\pm$0.1 &  765$\pm$20  &    7.8$\pm$ 0.1 \\
 3 &  & 0.93$\pm$0.01 &   84.4   &    5.4$\pm$0.2 &    6.2$\pm$0.2 &   14.7$\pm$ 0.2 \\
 4 &  & 0.29$\pm$0.01 &  114.3   &    7.2$\pm$0.3 &    4.6$\pm$0.2 &   13.9$\pm$ 0.1 \\
\hline
1 &100 -- 200 (BAT)   & 0.31$\pm$0.02 &   65.0   &   18.1$\pm$1.6 &   23$\pm$7 &    4.5$\pm$0.1 \\
 2 &  & 0.31$\pm$0.02 &   70.5   &    7.9$\pm$0.5 &  900$\pm$50 &    7.2$\pm$0.4 \\
 3 &  & 0.19$\pm$0.01 &   84.4   &    5.2$\pm$0.4 &    8.5$\pm$1.0 &   13.8$\pm$0.4 \\
 4 &  & 0.04$\pm$0.003 & 114.3   &    6.1$\pm$1.1 &    6.4$\pm$1.5 &   12.9$\pm$0.4 \\
\hline
\end{tabular}
\end{table}

\begin{table}
\label{table}
\centering
\linespread{0.02}
\caption{Results of cross-correlation.}
\begin{tabular}{llll}
\hline
Division No. &energy bands & mean energy (keV)& Time lag (s) \\
\hline
Part1 (T$_0$ to T$_0$+50) &15-25 vs 25-50 keV &35.45&  -2.0$\pm$0.07 \\
&15-25 vs 50-100 keV &68.07&  -3.85$\pm$0.10 \\
&15-25  vs 100-200 keV &123.73&  -7.00$\pm$0.14\\
\hline
Part2 (T$_0$+50 to T$_0$+77) &15-25 vs 25-50 keV &35.45&  -0.424$\pm$0.026 \\
&15-25 vs 50-100 keV &68.07&  -0.721$\pm$0.030\\
&15-25  vs 100-200 keV &123.73&  -1.050$\pm$0.035\\
\hline
Part3 (T$_0$+77 to T$_0$+100) &15-25 vs 25-50 keV &35.45&  -0.070$\pm$0.04 \\
&15-25 vs 50-100 keV &68.07&  -0.010$\pm$0.03\\
&15-25  vs 100-200 keV &123.73&  -0.210$\pm$0.033\\
\hline
Part4 (T$_0$+100 to T$_0$+180) &15-25 vs 25-50 keV &35.45&  -0.655$\pm$0.030 \\
&15-25 vs 50-100 keV &68.07&  -1.349$\pm$0.040\\
&15-25  vs 100-200 keV &123.73&  -1.775$\pm$0.052\\
\hline
Total (T$_0$ to T$_0$+180) & 15-25 vs 25-50 keV &35.45&  -0.495$\pm$0.026 \\
&15-25 vs 50-100 keV &68.07&  -0.872$\pm$0.028 \\
&15-25  vs 100-200 keV &123.73&  -1.310$\pm$0.035\\
\hline
\end{tabular}
\end{table}

\begin{table}[h]
\label{table}
\centering
\linespread{0.02}
\caption{Best fit spectral parameters from a combined fit to  BAT and $RT$-2 spectra along with timing parameters}
\begin{tabular}{llllllll}
\hline
Part & $\alpha$  & $\beta$ & $E_0$ & $\chi^2$/dof & $E_p$ & $\xi$   &   $d_i$ \\
     &          &        & (keV) & & (keV) & ($RT$-2 \& BAT) & \\
\hline
Full & -1.40$\pm$0.02     & -2.5$^{+0.3}_{-0.5}$   & 273$\pm$31 & 56 / 101 & 164$\pm$24  &  -- & -0.64    \\
Part1 &-1.18$^{+0.13}_{-0.08}$     & $<$ -1.6            & 322$^{+176}_{-103}$ 
& 52 / 55 & 264$^{+209}_{-102}$ & --  & -3.71 \\
 (precursor) & & & & &  \\
Part2 &-1.23$\pm$0.05         & $<$ -2.1            & 322$^{+79}_{-54}$
& 56/ 55 & 248$^{+81}_{-55}$ & 0.18$\pm$0.03 \&     & -0.50 \\
(pulse 1 \& 2) & & & & &  & 0.07$\pm$0.03  &  \\
Part3  & -1.39$\pm$0.03     & -2.4$\pm$0.2     &211$^{+20}_{-11}$
 & 55/ 55 & 129$^{+19}_{-13}$ & 0.14$\pm$0.02  & -0.13   \\
(pulse 3)  & & & & &  \\
Part4  &-1.70$^{+0.07}_{-0.10}$       &-2.8$^{+0.2}_{-1.1}$    & 111$^{+20}_{-14}$ 
& 37 / 55 & 33$^{+15}_{-14}$ &  0.05$\pm$0.01 & -0.92  \\
(pulse 4) & & & & &  \\
\hline
\hline
\end{tabular}
\end{table}

\end{document}